\documentclass[journal=jacsat,manuscript=article]{achemso}
\setkeys{acs}{articletitle=true}

\usepackage[version=3]{mhchem} % Formula subscripts using \ce{}
\usepackage[T1]{fontenc}       % Use modern font encodings
\usepackage{float}
\usepackage{graphicx}
\usepackage{subfig}
\usepackage{epsfig}
\usepackage{epstopdf}
\usepackage{enumerate}
\usepackage{physics}
\usepackage{longtable}
\usepackage{amssymb}
\usepackage{color}
\usepackage{colortbl}

\usepackage{xr-hyper}
\usepackage{hyperref}

\SectionNumbersOn
\definecolor{hellgrau}{rgb}{0.90,0.90,0.90}
\newcommand{\gs}{\gamma_{s}}
\newcommand{\RNum}[1]{\uppercase\expandafter{\romannumeral #1\relax}}
\newcommand{\rNum}[1]{\lowercase\expandafter{\romannumeral #1\relax}}

\author{Davinder Singh}
\email{davinder.mand@utoronto.ca}
\affiliation{Chemical Physics Theory Group, Department of Chemistry, and Center for Quantum
Information and Quantum Control, University of Toronto, Toronto, Ontario M5S 3H6, Canada}
\author{Paul Brumer}
\email{paul.brumer@utoronto.ca}
\affiliation{Chemical Physics Theory Group, Department of Chemistry, and Center for Quantum
Information and Quantum Control, University of Toronto, Toronto, Ontario M5S 3H6, Canada}

\title{%Coherent Control of Flux in a Donor-Acceptor Pair \\
Coherent Control of Energy Transport at Room Temperature in a Noisy Bath}


\begin{document}

%%%%%%%%%%%%%%%%%%%%%%%%%%%%%%%%%%%%%%%%%%%%%%%%%%%%%%%%%%%%%%%%%%%%%

\begin{tocentry}
{\centering
\includegraphics[width = 3.0 in]{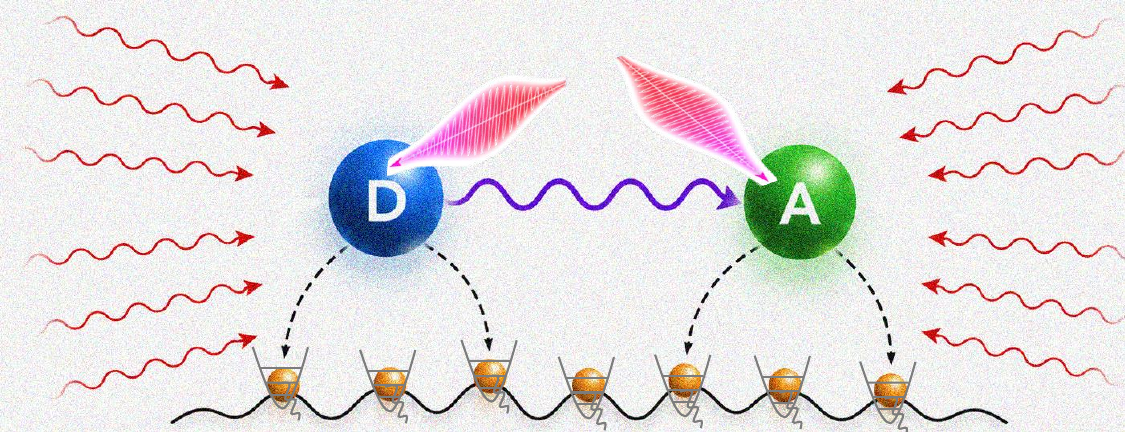}
\par
}
\end{tocentry}

%%%%%%%%%%%%%%%%%%%%%%%%%%%%%%%%%%%%%%%%%%%%%%%%%%%%%%%%%%%%%%%%%%%%%
\begin{abstract}
Coherent control of energy transport in a non-equilibrium steady-state (NESS) in a reaction-center-connected donor–acceptor pair is proposed. 
The pigments are considered to be continuously interacting with incoherent radiation and a phonon bath while being driven by phase-controlled coherent fields. 
Coherent excitation of the donor-acceptor pair is shown to induce interference between excitation pathways, resulting in phase dependent modulation of the flux.
As a consequence one can enhance or suppress energy transfer via interference, e.g. an optical energy switch.
The persistence of such interference enables coherent control at a NESS in dissipative regime suggests an extension of the operational scope of quantum control from traditional transient domain with low dissiaption to noisy environment NESS at room temperature.

\end{abstract}

%%%%%%%%%%%%%%%%%%%%%%%%%%%%%%%%%%%%%%%%%%%%%%%%%%%%%%%%%%%%
%\section{Introduction}

$Introduction-$Coherent control provides a methodology by which quantum interference effects between multiple initial pathways are used to enhance or minimize processes \cite{Quantum_Brumer_2011}, such as product cross sections in photodissociation or chemical reactions.
This interference has been explored in chemistry \cite{rouse_light_2024} and in systems like atomic vapours for optical quantum memory \cite{electromagnetically_harris_1997,PhysRevLett.81.5932,PhysRevLett.98.060502,singh_fundamental_2023,PhysRevLett.98.190503,ma_optical_2017} or trapped ions \cite{cirac_quantum_1995}, where the approach preserves coherent superpositions during the evolution of the quantum state.
The vast majority of formal, computational and experimental implementations of this approach have been applied to transient dynamics, where coherent evolution survives until the desired target is reached.
This often assumes minimizing any decoherence effects encountered during the dynamics in order to maintain the phase relationship between interfering pathways.
Strong decoherence is expected to reduce the extent of coherent control.

Here we consider a minimal model of biophysical systems that are driven by incident light, such as a donor-acceptor system in photosynthetic complexes. Interestingly, short time coherence can be generated in such complex systems in complex environments, e.g., via pulsed laser excitation of photosynthetic molecules like Fenna-Matthews-Olson (FMO) complex \cite{engel_evidence_2007,brixner_two-dimensional_2005,ishizaki_unified_2009,Pachon,singh_importance_2020}, phycocyanin 645 (PC645) \cite{mirkovic_ultrafast_2007,singh_coherent_2021}, as well as laser excited retinal in rhodopsin \cite{johnson_primary_2017,singh_machine_2024}, the first step in vision. In these cases the pulsed laser bandwidth creates the coherent superposition of states that subsequently evolves in time. The implication is that coherent control can be successful in these type of systems on timescales shorter than decoherence that destroy coherence.

However, it is important to note that while pulsed laser experiments provide valuable insights into the system Hamiltonian and its interaction with an environment, they are not the operational reality of many processes like natural light harvesting. 
Under in-situ conditions, pigment-protein complexes (PPCs) do not operate with coherent pulses, but operate with the incoherent excitation (e.g. natural light) of consistent intensity over molecular timescales\cite{kassal_does_2013}.
The realistic physical picture of natural light-harvesting involves continuous injection and migration of excitation energy in the system \cite{grinev_realistic_2015,dodin_coherent_2016,brumer_shedding_2018}. 
Hence, coherent control might well be achieved by affecting, via coherent interfering processes, the flux by modifying the associated off-diagonal density matrix elements. As shown in this paper, this successful approach constitutes a new direction for coherent control applications.

As an explicit example, we consider energy migration in a minimal model photosynthetic complex in the non-equilibrium steady state (NESS), a reaction center connected donor-acceptor pair in the presence of incoherent pumping and thermal phonons.
The controls are continuous-wave coherent driving fields and we observe that in a noisy environment at NESS, the magnitude of the flux can be controlled by coherent driving.
That is, in the NESS, the coherent phase relation between excited states can be protected by phase locked coherent beams and Coulomb coupling, even in the presence of environmental noise.
These findings support the extension of coherent control to noise-dominated systems at NESS, with implications for, e.g., artificial light-harvesting design and optical energy migration switches.

\begin{figure}[ht]
\includegraphics[width=0.5\columnwidth]{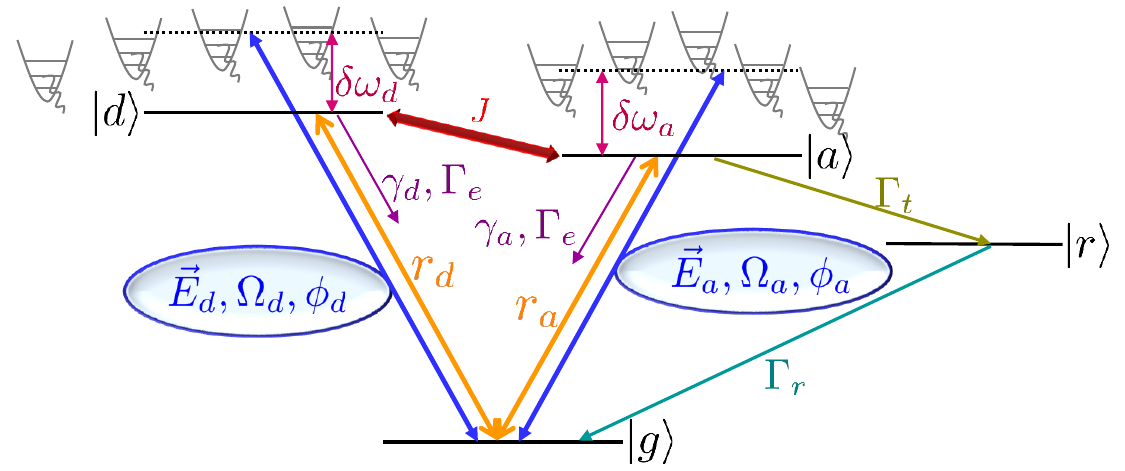}
\caption{Schematic of a reaction center connected donor-acceptor pair (with quantum states, common ground state $\ket{g}$, excited donor (acceptor) state $\ket{d}$ ($\ket{a}$), and reaction center state $\ket{r}$) in the presence of monochromatic control fields $\vec{E}_{m}$ (where $m = d,a$). Here, $\delta\omega_{m} = \omega_L - \omega_{mg}$ is the detuning of the monochromatic field, $r_m$ illustrates the incoherent pumping rate, $\gamma_m$ denotes the spontaneous decay rate, and $\Gamma_{t}(\Gamma_{r})$ represents the trapping (recycling) rate.  
}
\label{Levels}
\end{figure}

%\section{Theoretical model}
$Model-$Consider a donor-acceptor pair connected to the reaction center composed of quantum states $\ket{g}$, $\ket{d}$, $\ket{a}$ and $\ket{r}$ that are immersed in a bosonic bath.
The system interacts with incoherent light in addition to coherent control fields [see Fig.\ref{Levels}].
If interest is in model photosynthesis probed with ultrafast lasers, the process would involve a sequence of events where the absorption of light in a complex network of optically active molecules triggers excitation \cite{van_amerongen_photosynthetic_2000,blankenship_molecular_2002}. This excitation then migrates to the reaction center, leading to charge separation and the later dark stages of the process.
Here, however, we examine the more physically natural scenario where system excitation is carried out with continuous incoherent light, establishing a NESS.

The monochromatic field with polarization $\hat{e}_m$ ($m= d$, $a$) induces the excitations of $\ket{g}\rightarrow\ket{m}$. 
The incident monochromatic field is expressed as $\vec{E}_m(t) = \hat{e}_m\xi_m e^{-i\left(\omega_L t - \phi_m\right)} + h.c.$, where $\xi_m$ denotes the amplitude of the field with angular frequency $\omega_L$, and $\phi_m$ is the phase.
The transition dipole operator is given by 
$\vec{d} = \vec{d}_d+\vec{d}_a=\left(\vec{d}_{gd}\ket{g}\bra{d} + \vec{d}_{dg}\ket{d}\bra{g}\right) + \left(\vec{d}_{ga}\ket{g}\bra{a} + \vec{d}_{ag}\ket{a}\bra{g}\right)$ with the dipole matrix elements, $\vec{d}_{mn}$.
In the monochromatic field interaction (i.e., $ - \vec{d}_d.\vec{E}_{d} - \vec{d}_a.\vec{E}_{a}$), terms violating energy conservation are neglected within the dipole approximation \cite{scully_zubairy_1997}.
The system is surrounded by an incoherent radiation field that acts as a thermally equilibrated reservoir of harmonic oscillators that injects energy into the system. 
The excited states interact with a phonon bath maintained at temperature $T_b$, and the state $\ket{a}$ transfers the excitation energy to the reaction center $\ket{r}$, which ultimately connects to the ground state for the process's completion. 
The total Hamiltonian in the presence of all the interactions is expressed as \citep{jung_energy_2020,scully_zubairy_1997,Carmichael}
\begin{align}
&\hat{H}=\hat{H}_S+\hat{H}_{SE} + \hat{H}_R+ \hat{H}_{SR} + \hat{H}_B+ \hat{H}_{SB},
\label{Hamiltonian}
\end{align}
where
\begin{align}
&\hat{H}_S = \hbar\left( \omega_d\ket{d}\bra{d} + \omega_a\ket{a}\bra{a} + J \left( \ket{d}\bra{a} + \ket{a}\bra{d} \right) \right), \nonumber\\
&\hat{H}_{SE} = - \hbar\Omega_d \left(  e^{-i\left( \omega_Lt + \phi_d \right)}\ket{d}\bra{g} + e^{i\left( \omega_Lt + \phi_d \right)}\ket{g}\bra{d} \right) - \hbar\Omega_a \left(  e^{-i\left( \omega_Lt + \phi_a \right)}\ket{a}\bra{g} + e^{i\left( \omega_Lt + \phi_a \right)}\ket{g}\bra{a} \right), \nonumber\\
%-\hbar\left( Ge^{-i\omega_c t}\ket{1}\bra{2} + Ge^{i\omega_c t}\ket{2}\bra{1}+ge^{-i\omega_p t}\ket{1}\bra{3} + ge^{i\omega_p t}\ket{3}\bra{1} \right) \nonumber\\
&\hat{H}_R = \sum_{\textbf{k},\lambda} \hbar\omega_{\textbf{k},\lambda}r_{\textbf{k},\lambda}^{\dagger}r_{\textbf{k},\lambda}, \nonumber\\
&\hat{H}_{SR} = \sum\limits_{m=d,a} \sum_{\textbf{k},\lambda}\hbar\left[\left(g_{\textbf{k},\lambda}^{\ast}\right)^{m}r_{\textbf{k},\lambda}^{\dagger}\ket{g}\bra{m} + \left(g_{\textbf{k},\lambda}\right)^{m}r_{\textbf{k},\lambda}\ket{m}\bra{g} \right], \nonumber\\
&\hat{H}_B = \sum\limits_{l_{m=d,a}}  \hbar\omega_{l_m}b_{l_m}^{\dagger}b_{l_m}, \nonumber\\
&\hat{H}_{SB} = \sum\limits_{m=d,a} \ket{m}\bra{m} \sum\limits_{l_{m}} \hbar g_{l_m} \left( b_{l_m}^{\dagger} + b_{l_m} \right).
\label{Hamiltonian1}
\end{align}
Here, $\omega_m$ denotes the transition frequency of the $\ket{m}$th state, and $J$ is the dipole-dipole coupling. The monochromatic control field induces a coherent pumping rate $\Omega_m = d_{gm}.\xi_m/\hbar$. Moreover, $\omega_L$ is the frequency and $\phi_m$ is the phase of the monochromatic field. Note that $\Omega_m$ contains the polarization direction and $\xi_m$ is the field strength. In the radiation field interaction $H_{SR}$, the summation extends over the modes of the incident field with wavevector $\textbf{k}$ and polarization $\lambda$. 
The symbols, $r_{\textbf{k},\lambda}^{\dagger}$ and $r_{\textbf{k},\lambda}$ denote the creation and annihilation operators of the harmonic oscillators of angular frequency $\omega_k$ constituting the radiation reservoir. The dipole coupling constant, $\left(g_{\textbf{k},\lambda}\right)^m \equiv -i\sqrt{\omega_k/2\hbar\varepsilon_0V}\hat{e}_{\textbf{k},\lambda}\cdot\vec{d}_{gm}$ for $m \in \{d,a\}$, 
contains the information of polarization $\hat{e}_{\textbf{k},\lambda}$, quantization volume $V$ and vacuum permittivity $\varepsilon_0$. Further, $b_{l_m}^{\dagger}$ and $b_{l_m}$ represent the creation and annihilation operators of the bath harmonic oscillators of angular frequency $\omega_{l_m}$, and $g_{l_m}$ describes the system-bath coupling.

The matrix formalism describing the evolution of the system density matrix $\hat{\rho}(t)$ amid all the interactions is given by
%\begin{widetext}
\begin{align}
\frac{d\hat{\rho}(t)}{dt} =& - \frac{i}{\hbar}[\hat{H}_{S} +\hat{H}_{SE}, \rho] + \mathcal{L}_{SB} + \mathcal{L}_{SR} + \mathcal{L}_{e} + \mathcal{L}_{t} + \mathcal{L}_{r} .
\label{master}
\end{align}
%\end{widetext} 
A detailed discussion of the system-bath $\mathcal{L}_{SB}$ [given in subsection S$1.1$], the system-radiation field $\mathcal{L}_{SR}$ [given in subsection S$1.2$], exciton recombination $\mathcal{L}_{e}$ [given in subsection S$1.3$], energy transfer to trap state $\mathcal{L}_{t}$ [given in subsection S$1.3$], and $\mathcal{L}_{r}$ [given in subsection S$1.3$] is provided in the supporting information (SI).
By transforming the density matrix into the rotating frame, that is\cite{singh_origin_2021}, $\hat{\rho} \rightarrow \tilde{\rho}$, the coherence-flux relation can be expressed as \cite{yang_steady-state_2020,singh_origin_2021}, 
\begin{align}
\mathcal{F}\propto \tilde{\rho}_{ad}^I,
\label{flux_coh}
\end{align}
where the $\mathcal{F}$ is the energy flux between donor and acceptor molecule and $\tilde{\rho}_{ad}^I$ is imaginary part of the coherence between states $\ket{d}$ and $\ket{a}$. For a discussion of the range of validity of Eq.(\ref{flux_coh}) see \cite{jacobs_2026}. 
The dynamics of donor-acceptor coherence can be written as
\begin{align}
\frac{d\tilde{\rho}_{ad}}{dt}
={}&
-\left(
\Gamma_{ad}^{\mathrm{coh}}
-i\Delta_{ad}^{\mathrm{eff}}
\right)
\tilde{\rho}_{ad} -
\left(
\kappa+iJ
\right)
\left(\tilde{\rho}_{dd}-\tilde{\rho}_{aa}\right) -
\left(
i\mu+C_{ad}^{\downarrow}
\right)
\left(\tilde{\rho}_{dd}+\tilde{\rho}_{aa}\right)
+
C_{ad}^{\uparrow}
\tilde{\rho}_{gg}
\nonumber\\
&-
i\Omega_d
e^{i\phi_d}
\tilde{\rho}_{ag}
+
i\Omega_a
e^{-i\phi_a}
\tilde{\rho}_{dg}^{\ast}.
\label{dynamical_equation2}
\end{align}
Eq.(\ref{dynamical_equation2}) shows the different physical mechanisms governing the donor-acceptor coherence. The first term of Eq.(\ref{dynamical_equation2}) describes coherent phase evolution with bath-renormalized detuning $\Delta_{ad}^{\mathrm{eff}}\equiv{} \delta\omega_a-\delta\omega_d - \left( \mu_{4,d}-\mu_{3,d} +\mu_{3,a}-\mu_{4,a} \right)$ along with phase damping at $\Gamma_{ad}^{\mathrm{coh}}\equiv{} \kappa_{3,d}+\kappa_{3,a} +\kappa_{4,d}+\kappa_{4,a} + \frac{1}{2} \left( r_d+r_a+\gamma_d+\gamma_a \right) +2\Gamma_e+\Gamma_t$.
Here, $\delta\omega_{m} = (\omega_L - \omega_{m})$ is the detuning of the monochromatic field from the transition $\ket{g} \rightarrow \ket{m}$, $ \kappa_{j,m} $ and $\mu_{j,m} $ (with $j = 1,2,3,4$) are the phonon bath induced de-phasing or relaxation rate [see  S$1.1$], $r_m $ is the incoherent pumping rate, $\gamma_m$ is spontaneous decay rate [see  S$1.2$], $\Gamma_{e}$ is exciton recombination rate, and $\Gamma_{t}$ is trapping rate from the acceptor to the reaction center [see  S$1.3$].
The second term of Eq.(\ref{dynamical_equation2}) illustrates the population imbalance effects the coherence via Coulomb coupling and phonon bath induced rate $\kappa \equiv \left( \kappa_{2,d}^{R} + \kappa_{2,a}^{R}  \right) +i \left( \kappa_{2,d}^{I} + \kappa_{2,a}^{I} \right)$, the third term connects the total singly excited population to coherence via imaginary part of the phonon-bath induced rate $\mu \equiv \left(\mu_{2,d}^{R}+\mu_{2,a}^{R}\right)+i\left(\mu_{2,d}^{I}+\mu_{2,a}^{I}\right)$ and thermal radiation induced transition rate $C_{ad}^{\downarrow}\equiv\frac{\textbf{p}}{2}\left(\sqrt{r_dr_a}+\sqrt{\gamma_d\gamma_a}\right)$ (here, $\textbf{p}$ is the alignment factor [see  S$1.2$]), the fourth term connects the coherence to ground state population through thermal radiation induced transition rate $C_{ad}^{\uparrow}\equiv\textbf{p}\sqrt{r_dr_a}$, and the final two terms describe the effect of the ground-to-excited state coherences generated by the two phase-controlled monochromatic fields on the donor-acceptor coherence. 
The NESS is given by the solution to $\frac{d\tilde{\rho}(t)}{dt} =0$.
Note that the exact analytical solutions for the NESS are too lengthy and complicated to allow direct physical insight (see subsection S$2$ of SI).
However, with some particular assumptions, basic physical insights can be obtained (see the subsection S$2$ of SI).
Despite the algebraic complexity, numerical simulations based on the complete analytical expressions are obtained.
In the simulations, the spontaneous decay rate $\gamma_m$ is considered as a function of transition frequency $\omega_m$ such that $\gamma_m = \left(1/4\pi\epsilon_0 \right)\left(4\omega_m^3 d_{mg}^2/3\hbar c^3 \right)$\citep{Carmichael}. We use $1/4\pi\epsilon_0 = 9\times10^9$ $Nm^2C^{-2}$, the transition dipole moment $d_{mg}^2 = 37.1\mathcal{D}^2$ with $\mathcal{D}=3.335\times 10^{-30}$ $Cm$, and $c = 3\times 10^8$ $ms^{-1}$. 
The incoherent pumping rate $r_m$ characterizes the number of incident photons per second from an incoherent source. To calculate $r_m$, the absorption cross-section $\sigma_{ab}$ is calculated using the molar extinction coefficient \cite{blankenship_molecular_2002}, where for, e.g., phycobilisomes, the molar extinction coefficient is observed to be $\sim 2.4\times10^6$ mol$^{-1}$cm$^{-1}$ with an absorption maxima around $652$ nm (corresponding to $15340$ cm$^{-1}$) \cite{glazer_light_1985}. The incoherent intensity ($I_{in}$) can be used to calculate the incoherent pumping rate such that $r_m = \sigma_{ab} I_{in}/\hbar \omega_m$.
Similarly, the coherent pumping rate is a function of coherent source intensity ($I_{ch}$), that is, $\Omega_m = \sqrt{3\gamma_m\lambda_m^3 I_{ch}/2\pi h c}$ \cite{singh_fundamental_2023}.
Other parameters used are: $\omega_d = \omega_a = 2\pi \times 15340$ cm$^{-1}$, peak phonon bath frequency $\gamma_d^b = \gamma_a^b = 2\pi \times 100$ cm$^{-1}$, reorganization energy $\lambda_d^b = \lambda_a^b = 2\pi \times 10$ cm$^{-1}$, phonon bath temperature $T_b = 300$ $K$, alignment factor $\textbf{p} = .2$, exciton recombination rate $\Gamma_e = 2\pi \times 1/300$ cm$^{-1}$, trapping rate $\Gamma_t = 2\pi \times 33.34$ cm$^{-1}$, recycle rate $\Gamma_r = 2\pi \times 1/300$ cm$^{-1}$, incoherent intensity\cite{kruse_photosynthesis_2005} $I_{in} = 413$ watts m$^{-2}$, coherent intensity $I_{ch} = 100$ watts m$^{-2}$ (unless specified). 

\begin{figure*}[ht]
\includegraphics[width=0.87\columnwidth]{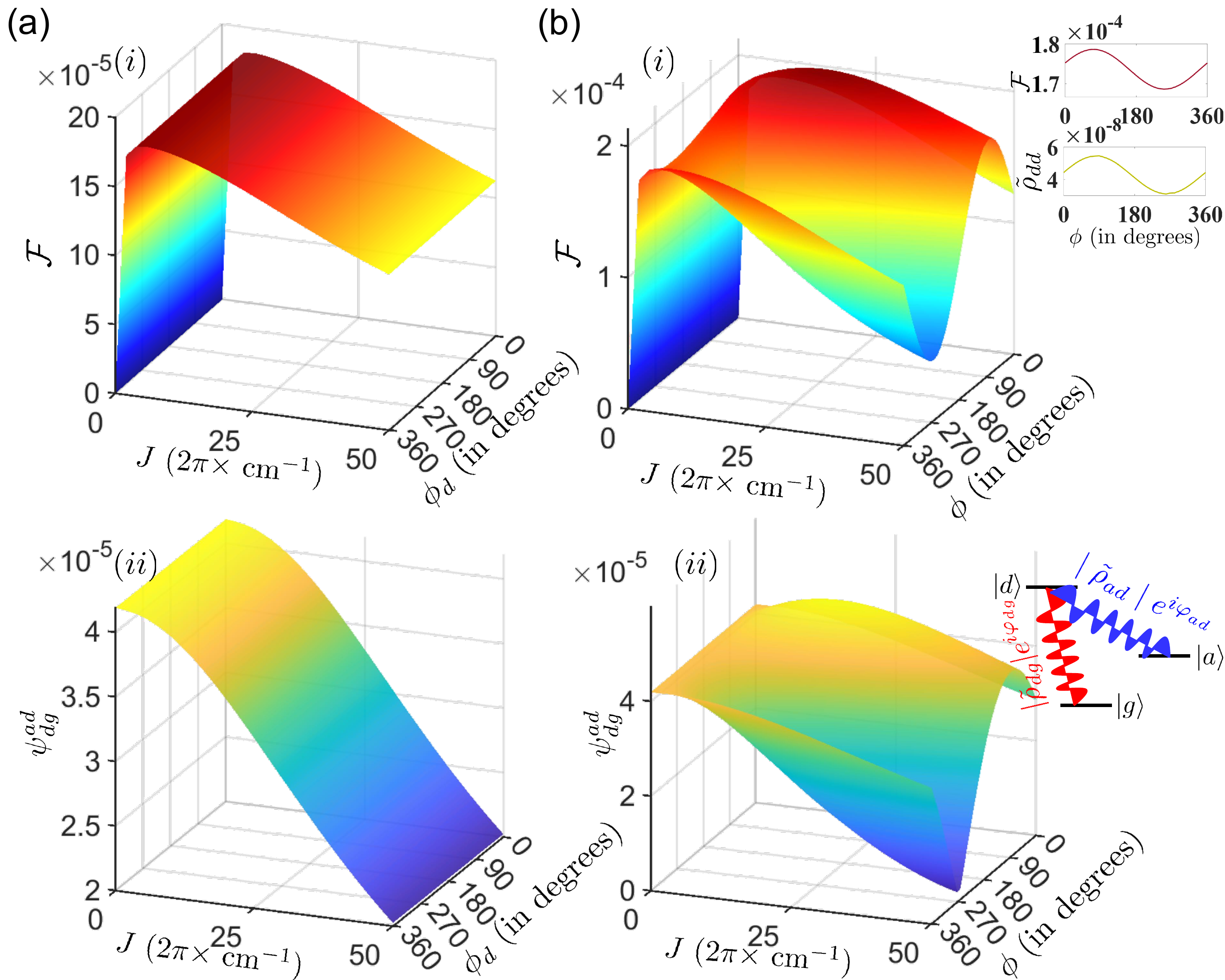}
\caption{Illustration of $(\rNum{1})$ flux and $(\rNum{2})$ interference pattern as a function of the relative phase between laser beams and the Coulomb coupling $J$ (a) in the absence [or presence (b)] of $\ket{g}\rightarrow\ket{a}$ coherent beam. The coherent beam for transition $\ket{g}\rightarrow\ket{d}$ is on in both cases. Further, inset for b$(\rNum{1})$ shows the dependence of the flux (upper panel) on the donor population (lower panel) for $J = 2\pi\times 2$ cm$^{-1}$. Inset: for b$(\rNum{2})$ depicts the interference pathways when both the laser beams are switched on. Note that for computational purposes, NESS solutions and hence the flux $\mathcal{F}$ is rescaled with respect to a normalization parameter $\gamma = 2\pi \times10^{-3}$ cm$^{-1}$.
}
\label{Flux_G_phi}
\end{figure*}

%\section{Results and Discussions}

\textit{Effect of relative phase among coherent beams}$-$ To explore the possibility of laser-induced phase control in a noisy environment, we varied the relative phase $\phi$ $(= \phi_d - \phi_a)$ between coherent beams.
The phase dependence of flux is then explored using the full analytical expression, maintaining quantum interference and its effects in a noisy environment.
Some analytic insight is afforded by examining a resonant case with negligible Coulomb coupling and negligible trapping rate. The NESS values for the donor-acceptor coherence for this case can then be expressed as
\begin{align}
(\tilde{\rho}_{ad}^R)^{ss} & = \mathcal{D}_s^{-1} (\Omega_a \Omega _d \left(4 \Omega_a^2 \gamma_d+4 \gamma_a \Omega _d^2+\gamma_a \gamma_d \left(\gamma_a+\gamma_d\right)\right) \cos \left(\phi _d-\phi _a\right)) \nonumber\\
(\tilde{\rho}_{ad}^I)^{ss} & = \mathcal{D}_s^{-1} (4 \Omega_a \Omega _d \left(4 \Omega_a^2 \gamma_d+4 \gamma_a \Omega _d^2+\gamma_a \gamma_d \left(\gamma_a+\gamma_d\right)\right) \sin \left(\phi _d-\phi _a\right)). \nonumber\\
\label{phase_dep}
\end{align}
Here, $\mathcal{D}_s = 32 \gamma_a \Omega _d^4+4 \Omega _d^2 \left(8 \Omega_a^2 \left(\gamma_a+\gamma_d\right)+\gamma_a \left(2 \gamma_a^2+2 \gamma_a \gamma_d+\gamma_d^2\right)\right)+\gamma_d \left(\gamma_a^2+8 \Omega_a^2\right) \left(\gamma_d \left(\gamma_a+\gamma_d\right)+4 \Omega_a^2\right)$. % and $\phi = \phi_d - \phi_a$ is the relative phase.
 Eq.(\ref{phase_dep}) shows the direct dependence of the NESS donor-acceptor flux $\mathcal{F}$, through $(\tilde{\rho}_{ad}^I)^{ss}$, on the relative phase.

To explore the interference consider the flux dependence evident in, for example, the sum $\psi_{dg}^{ad} = \mid \tilde{\rho}_{dg} \mid e^{i\varphi_{dg}} + \mid \tilde{\rho}_{ad} \mid e^{i\varphi_{ad}}$ [see Fig.\ref{Flux_G_phi}b($\rNum{2}$) inset]. Here, $\mid \tilde{\rho}_{mn} \mid^2 = \left( \tilde{\rho}_{mn}^R \right)^2 + \left( \tilde{\rho}_{mn}^I \right)^2$ and $\varphi_{mn} = \tan^{-1}\left( \tilde{\rho}_{mn}^I/\tilde{\rho}_{mn}^R \right)$. 
Initially only one laser beam (for the transition $\ket{g}\rightarrow\ket{d}$) is considered in the simulation such that $\Omega_d \neq 0$ and $\Omega_a \simeq 0$. 
Clearly, in this case, no interference pattern [see Fig.\ref{Flux_G_phi}a($\rNum{2}$)] is observed with changes in the phase $\phi_d$. In addition, the flux [see Fig.\ref{Flux_G_phi}a($\rNum{1}$)] stays almost constant as a function of $\phi_d$.
On the other hand, when both the beams are on (such that $\Omega_d = \Omega_a \neq 0 $), the interference pattern appears as a function of the relative phase $\phi$ [see Fig.\ref{Flux_G_phi}b($\rNum{2}$)] due to interference between the two-pathways described by transitions $\ket{g}\leftrightarrow\ket{d}$ and $\ket{d}\leftrightarrow\ket{a}$ [see Fig.\ref{Flux_G_phi}b($\rNum{2}$) inset]. 
Thus, for a reaction-center connected donor-acceptor pair, the phase dependent laser-induced interference pattern survives even for intermediate bath de-phasing.
This interference pattern also affects the flux [see Fig.\ref{Flux_G_phi}b($\rNum{1}$)].
With the change in the relative phase, variations in the interference pattern govern the flux such that the maxima of the interference pattern corresponds to enhanced flow of energy between donor-acceptor molecule and vice-versa [see Fig.\ref{Flux_G_phi}(b)]. Moreover, the amplitude of the resultant wave $\psi_{dg}^{ad}$ and $\mathcal{F}$ is a non-monotonic function of $J$. [Eq.(S$18$) already shows the non-monotonic behaviour of the flux as a function of Coulomb coupling].
Essentially, too weak Coulomb coupling reduces the flow of energy, while strong Coulomb coupling, i.e., $J\gtrsim 2\pi\times 40$ cm$^{1}$ creates strong overlap among excited states which obstructs directional flow to the target site.
As with the coherence $\tilde{\rho}_{ad}$, the phase-induced interference affects the population ($\tilde{\rho}_{dd}$) [see Fig.\ref{Flux_G_phi}b($\rNum{1}$) inset] since the donor-acceptor pair population and coherence are coupled to one another in the NESS \cite{yang_steady-state_2020}.

\begin{figure*}[ht]
\includegraphics[width=1.0\columnwidth]{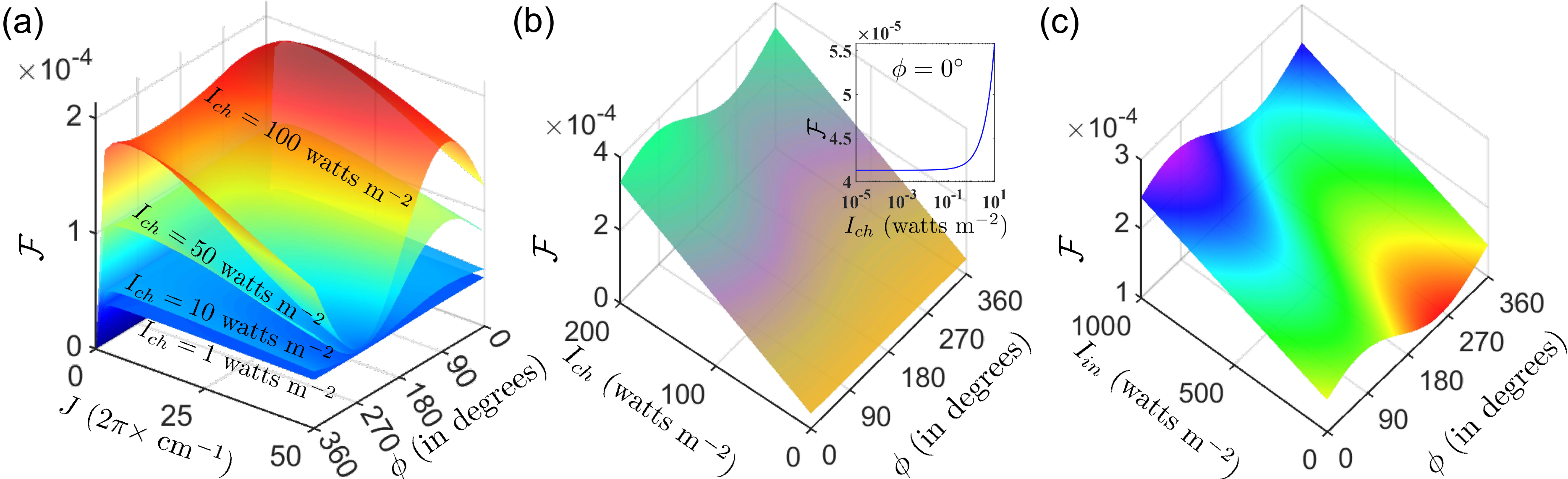}
\caption{(a) Flux as a function of $\phi$ and $J$ with different values of coherent source intensity. (b) Flux as a function of $\phi$ and coherent source intensity. (c) Flux as a function of $\phi$ and incoherent source intensity. Here, we have used $J = 2\pi \times 10$ cm$^{-1}$ in the simulation of (b)\&(c).
}
\label{Flux_phi}
\end{figure*}

Further, the flux dependence on different values of the coherent source intensity demonstrates loss of coherent control of the flux magnitude [see Fig.\ref{Flux_phi}(a)] when $I_{ch} = 1$ watt m$^{-2}$. 
That is, there is a requirement for a minimum intensity of the coherent light source for control over the flux amplitude. With $I_{ch} = 1$ watt m$^{-2}$, the flux injected via a coherent channel is about $6\%$ of incoherently injected flux [see Fig.\ref{Flux_phi}(b) inset], while with $I_{ch} = 10$ watts m$^{-2}$, the flux injected via a coherent channel is $\sim 35\%$ of incoherently injected flux. Moreover, at the minima of the coherently controlled flux (i.e., $J \sim 2\pi \times 50$ cm$^{-1}$ and $\phi\sim 250^\circ$) the only thing that survives is the incoherently injected flux. 
That is, quantum interference mainly controls the flux injected via a coherent source. 
Such a blockade of the flux via destructive interference supports the idea of flux gating/switching via phase, analogous to an optical transistor enabled by quantum effects in noisy settings. 
Recently, at cryogenic temperature, in the design of single-molecule transistors, the effect of quantum interference has been experimentally observed on the flux gating \cite{chen_quantum_2024}. 
Therefore, insights from this work suggest the possibility of design principles for a room-temperature (or colder) picosecond quantum switch, where flux is coherently controlled, with noise resilience due to the system's ability to preserve interference under dissipative conditions.

As the coherent intensity increases, so does the variation of flux minima and maxima with $\phi$ [see Fig.\ref{Flux_phi}(b)]. 
Apart from this quantum interference effect, the injection of flux is a linear function of both the coherent and incoherent pumping [see Fig.\ref{Flux_phi}(b) \& (c)]. As noted above, coherent pumping is stronger in injecting flux than is incoherent injection. For instance, the amplitude of the flux with $I_{in} = 413$ watts m$^{-2}$ is $\sim 0.4 \times 10^{-4}$, while with $I_{ch} = 50$ watts m$^{-2}$ is $\sim 1 \times 10^{-4}$ for $J = 2\pi \times 25$ cm$^{-1}$ and $\phi = 360^\circ$ [see Fig.\ref{Flux_phi}(a)]. 
Basically, coherent pumping leads to strong injection of flux compared to incoherent pumping since the latter affects population only, resulting in limited incoherent flux injection. 
Overall, our simulations illustrate that energy flux is not only governed by the excitation strength, but also that the flux amplitude can be coherently controlled between donor and acceptor pathways.

\begin{figure*}[ht]
\includegraphics[width=0.8\columnwidth]{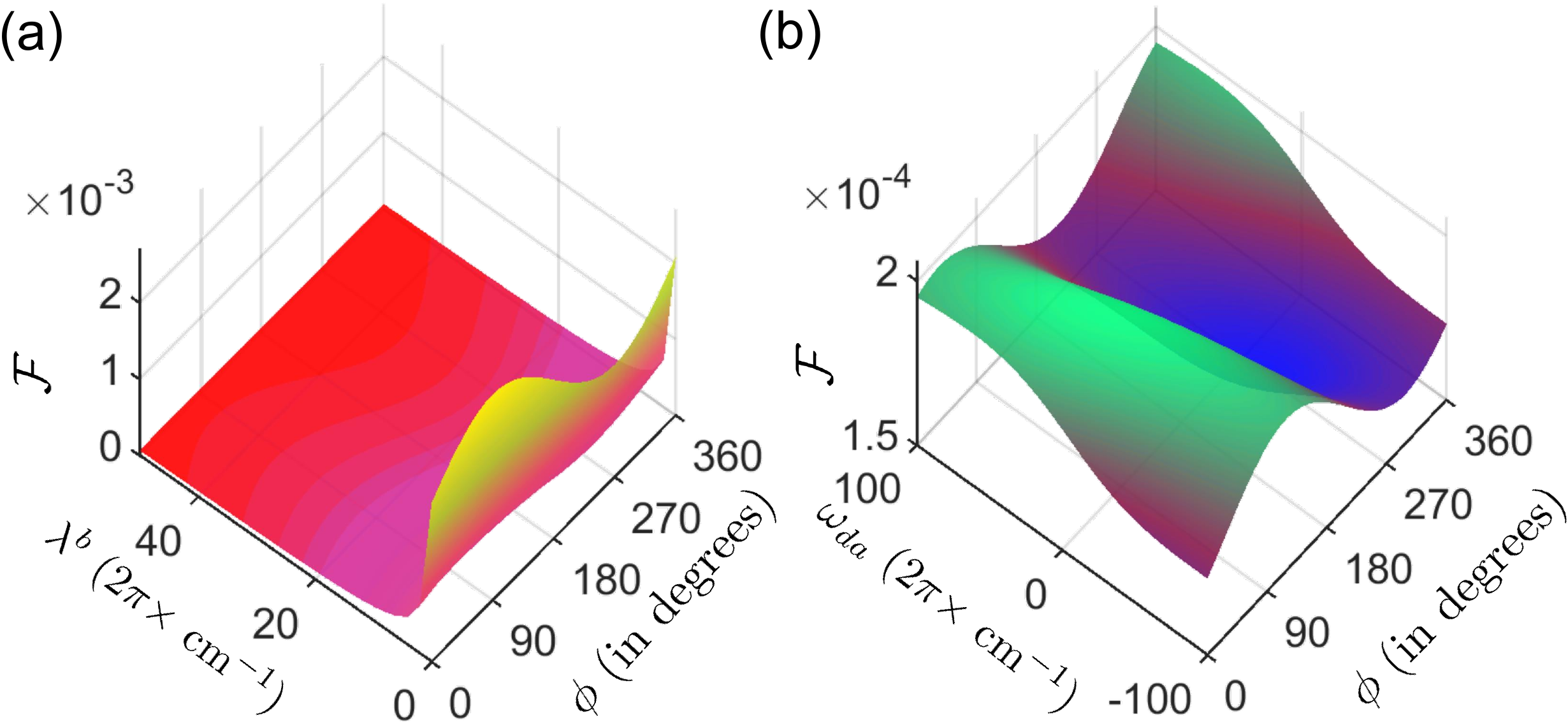}
\caption{(a) The effect of the system phonon bath coupling on flux for the different values of relative phase. (b) Evaluation of flux for the asymmetric donor-pair acceptor, i.e., $\delta\omega_{da} = \omega_d - \omega_a \neq 0$ as a function of $\phi$. In this case, two laser beams are considered such that laser beam for $\ket{g} \rightarrow\ket{d}$ transition remains resonant, while laser beam for $\ket{g} \rightarrow\ket{a}$ transition becomes off-resonant with $\delta\omega_{da} \neq 0$. Here, we have used $J = 2\pi \times 10$ cm$^{-1}$.
}
\label{Flux_phi_v1}
\end{figure*}

To analyze the competition between coherent control and environmental decoherence, we examine the effect of the system bath coupling on control [see Fig.\ref{Flux_phi_v1}(a)]. A strong phase dependence of flux is observed at low system phonon bath coupling with pronounced peaks and valleys indicating robust quantum interference effects. As expected, the enhancement in phonon bath dissipation-dephasing is detrimental to the flux amplitude.  
At strong system phonon bath coupling, almost complete suppression of quantum control is observed such that the flux becomes nearly phase-independent.
The persistence of control up to moderate system phonon bath coupling ($\sim \lambda^b \leq 2\pi \times 20$ cm$^{-1}$) suggests a potential design window for coherent control, where environmental effects are present but not fully detrimental.

In addition to the symmetric donor-acceptor pair, the asymmetric case is of interest and is characterized by the donor-acceptor energy gap $\delta\omega_{da} = \omega_d - \omega_a$. %is also analyzed by changing $\delta\omega_{da}$ . 
Even with non-zero $\delta\omega_{da}$, the flux follows a sinusoidal or quasi-sinusoidal profile as a function of $\phi$ [see Fig.\ref{Flux_phi_v1}(b)], indicating the persistence of interference in the presence of the energetic mismatch.
However, the maxima and minima are no longer symmetric around a particular value of $\phi$, as in the symmetric case. 
This variation reflects the loss of mirror symmetry between donor and acceptor energy levels.
The persistence of an interference pattern with non-zero $\delta\omega_{da}$ demonstrates the robustness of coherent control, even when perfect symmetry is not maintained.
This observation is particularly relevant for PPCs (for example FMO complex), where site energies are inherently disordered, suggesting that coherent control remains a viable mechanism for manipulating flux in realistic asymmetric systems.

%\section{Conclusions}

$Conclusions-$ 
In this study, a novel method is introduced to coherently control flux in a minimal model of systems such as photosynthetic pigments in a NESS. The model includes realistic bath dephasing mechanisms in a condensed phase with intermediate phononic couplings.
Numerical simulations show that, in a model donor-acceptor pair, 
the introduction of coherent interference by varying the relative phase of laser beams allows for a significant manipulation of the flux between the donor and acceptor. 
These results extend the scope of coherent control beyond the conventional realm of the transient domain, to be viewed as a tool for regulating steady-state transport in open quantum systems. 

\begin{acknowledgement}
We acknowledge financial support from NSERC Canada. We recognize the scientific guidance of Professor Jianshu Cao in this research field. His seminal contributions have been central to developments in Chemical Physics.
\end{acknowledgement}

\begin{suppinfo}

	The Supporting Information is available free of charge at
	\begin{itemize}
		\item Detailed description of the system-bath interaction, system-radiation bath, non-unitary part of dissipator; analytical form of the NESS flux.
	\end{itemize}
	
\end{suppinfo}

\newpage

\begin{center}
{\Large \textbf{\underline{Supporting Information}}}\\[0.5em]
{\Large Coherent Control of Energy Transport at Room Temperature in a Noisy Bath}
\end{center}

% Reset counters for Supplementary Materials
\setcounter{section}{0}
\setcounter{subsection}{0}
\setcounter{figure}{0}
\setcounter{equation}{0}

% Supplementary numbering
\renewcommand{\thesection}{S\arabic{section}}
\renewcommand{\thesubsection}{\thesection.\arabic{subsection}}
\renewcommand{\thefigure}{S\arabic{figure}}
\renewcommand{\theequation}{S\arabic{equation}}

\section{Dissipator for the master equation}
\label{sec:ms_eq}
\subsection{System-bath Liouvillian}
\label{subsec:System-bath part}

Under the Markovian approximation and weak coupling, the system-bath Liouvillian simplifies to \cite{Carmichael}
\begin{align}
\mathcal{L}_{SB}
= -\sum_{m\in\{d,a\}}\int_{0}^{\infty} dt\,
\Bigg(
&\alpha_m^{R}(t)\,
\commutator{P_m}{\commutator{P_m(-t)}{\rho}}
+i\alpha_m^{I}(t)\,
\commutator{P_m}{\anticommutator{P_m(-t)}{\rho}}
\Bigg),
\label{Eq_bath}
\end{align}
here, $\left[ ..\right]$ ($\{ ..\}$) represents the commutator (anti-commutator), $P_m \equiv \ket{m}\bra{m}$ such that $m\in\{d,a\}$ and $\alpha_{m}^R(t)$ and $\alpha_{m}^I(t)$ are the real and imaginary parts of the bath correlation function. %, responsible for the dephasing and dissipation.  
Their explicit expressions are given as follows by assuming the Lorentz-Drude form of spectral density $J_m(\omega)=\frac{2\lambda_m^b}{\pi}\frac{\gamma_m^b\omega}{(\gamma_m^b)^2+\omega^2}$, where $\lambda_m^b=\int_0^{\infty} \frac{J_m(\omega)}{\omega}d\omega$ is the reorganization energy and $\gamma_m^b$ involve the peak position and the width of the spectral density, which define the time scale of dephasing and dissipation to the phonon bath \cite{singh_survival_2021}:
\begin{align}
\alpha_m^R(t)=&\lambda_m^b\gamma_m^b\left[\cot{\left(\beta \gamma_m^b/2 \right)}e^{-\gamma_m^b t} + \frac{4}{\beta\gamma_m^b}\sum\limits_{q = 1}^{\infty} \frac{(\nu_{q}/\gamma_m^b)}{(\nu_{q}/\gamma_m^b)^{2} - 1} e^{-\nu_{q}t}\right], 
\end{align}
where $\nu_{q} = \frac{2\pi q}{\beta }$ corresponds to the bosonic Matsubara frequency. The imaginary part is  
\begin{align}
\alpha_m^I(t)=-\lambda_m^b\gamma_m^b e^{-\gamma_m^b t}. 
\end{align}
For weak system-bath coupling, small values of the re-organization energy are used. 

Further, the time dependence of the operator $\ket{m}\bra{m}$ can be written in the interaction picture as follows
\begin{equation}
\ket{m}\bra{m}(t) = e^{i\hat{H}_St}\ket{m}\bra{m}e^{-i\hat{H}_St}\;.
\label{sys_bath}
\end{equation}

The Baker-Campbell-Hausdorff formula with $\hat{H}_S$ allows one to rewrite $\ket{d}\bra{d}$ and $\ket{a}\bra{a}$ in the following form: 
\begin{align}
\ket{d}\bra{d}(t)& = \eta_1(t)\ket{d}\bra{a} - \eta_2(t)\ket{a}\bra{d} +  [\eta_3(t) + 1]\ket{d}\bra{d} -  \eta_3(t)\ket{a}\bra{a}, \nonumber\\
\ket{a}\bra{a}(t)& = \eta_2(t)\ket{a}\bra{d} - \eta_1(t)\ket{d}\bra{a} + [\eta_3(t) + 1] \ket{a}\bra{a} - \eta_3(t) \ket{d}\bra{d} . 
\label{sys_bath1}
\end{align}
here $\eta_1(t) = iJ\Omega_r^{-1} \sin(-\Omega_r t) - J(\omega_d - \omega_a)\Omega_r^{-2}[\cos(-\Omega_r t) - 1]$ and $\eta_2(t) = iJ\Omega_r^{-1} \sin(-\Omega_r t) + J(\omega_d - \omega_a)\Omega_r^{-2}[\cos(-\Omega_r t) - 1]$ are the coefficients describing population–coherence mixing. Moreover, the donor–acceptor population mixing is a function of $\eta_3(t) = 2J^2\Omega_r^{-2}[\cos(-\Omega_r t) - 1]$, and $\Omega_r = \sqrt{(\omega_{d} - \omega_{a})^{2}+ 4J^2}$ determines the frequency of coherent population exchange between donor and acceptor.

By using $P_d=\ket{d}\bra{d}, P_a=\ket{a}\bra{a}, T_d=\ket{d}\bra{a}, T_a=\ket{a}\bra{d}.$ such that $T_d^\dagger=T_a$ and $T_a^\dagger=T_d$, Eq.(\ref{Eq_bath}) can be expressed as
\begin{equation}
\mathcal L_{SB}
=
\mathcal L_d^{(R)}
+\mathcal L_a^{(R)}
+\mathcal L_d^{(I)}
+\mathcal L_a^{(I)},
\end{equation}
where
\begin{align*}
\mathcal L_d^{(R)}
={}&
\kappa_{1,d}
\left(
-T_d\rho+P_d\rho T_d+T_d\rho P_d
\right)
+
\kappa_{2,d}
\left(
\rho T_d^\dagger
-P_d\rho T_d^\dagger
-T_d^\dagger\rho P_d
\right)
\nonumber\\
&-
\kappa_{3,d}
\left(
\anticommutator{P_d}{\rho}
-2P_d\rho P_d
\right)
-
\kappa_{4,d}
\left(
P_d\rho P_a+P_a\rho P_d
\right),\\
\mathcal L_a^{(R)}
={}&
\kappa_{2,a}
\left(
-T_a\rho+P_a\rho T_a+T_a\rho P_a
\right)
+
\kappa_{1,a}
\left(
\rho T_a^\dagger
-P_a\rho T_a^\dagger
-T_a^\dagger\rho P_a
\right)
\nonumber\\
&-
\kappa_{3,a}
\left(
\anticommutator{P_a}{\rho}
-2P_a\rho P_a
\right)
-
\kappa_{4,a}
\left(
P_a\rho P_d+P_d\rho P_a
\right),\\
\mathcal L_d^{(I)}
={}&
-i\mu_{1,d}
\left(
T_d\rho+P_d\rho T_d-T_d\rho P_d
\right)
-
i\mu_{2,d}
\left(
\rho T_d^\dagger
-P_d\rho T_d^\dagger
+T_d^\dagger\rho P_d
\right)
\nonumber\\
&-
i\mu_{3,d}\commutator{P_d}{\rho}
+
i\mu_{4,d}
\left(
P_d\rho P_a-P_a\rho P_d
\right),\\
\mathcal L_a^{(I)}
={}&
-i\mu_{2,a}
\left(
T_a\rho+P_a\rho T_a-T_a\rho P_a
\right)
-
i\mu_{1,a}
\left(
\rho T_a^\dagger
-P_a\rho T_a^\dagger
+T_a^\dagger\rho P_a
\right)
\nonumber\\
&-
i\mu_{3,a}\commutator{P_a}{\rho}
+
i\mu_{4,a}
\left(
P_a\rho P_d-P_d\rho P_a
\right).
\end{align*}

The coefficients are
\begin{align*}
\kappa_{1,m}
&=
\int_0^\infty dt\,
\eta_1(-t)\alpha_m^R(t),
&
\kappa_{2,m}
&=
\int_0^\infty dt\,
\eta_2(-t)\alpha_m^R(t),
\\
\kappa_{3,m}
&=
\int_0^\infty dt\,
\left[1+\eta_3(-t)\right]\alpha_m^R(t),
&
\kappa_{4,m}
&=
\int_0^\infty dt\,
\eta_3(-t)\alpha_m^R(t),
\\
\mu_{1,m}
&=
\int_0^\infty dt\,
\eta_1(-t)\alpha_m^I(t),
&
\mu_{2,m}
&=
\int_0^\infty dt\,
\eta_2(-t)\alpha_m^I(t),
\\
\mu_{3,m}
&=
\int_0^\infty dt\,
\left[1+\eta_3(-t)\right]\alpha_m^I(t),
&
\mu_{4,m}
&=
\int_0^\infty dt\,
\eta_3(-t)\alpha_m^I(t).
\end{align*}
Here, the real part of the bath correlation dependent coefficients ($\kappa_{j,m}$) describe dissipative effects, including dephasing and relaxation, and the coefficients associated with the imaginary part of the bath correlation function ($\mu_{j,m}$) describe dispersive and Lamb-shift-like contributions.

\subsubsection{Redfield vs Lindblad approach}
\begin{figure*}[ht]
\includegraphics[width=0.7\columnwidth]{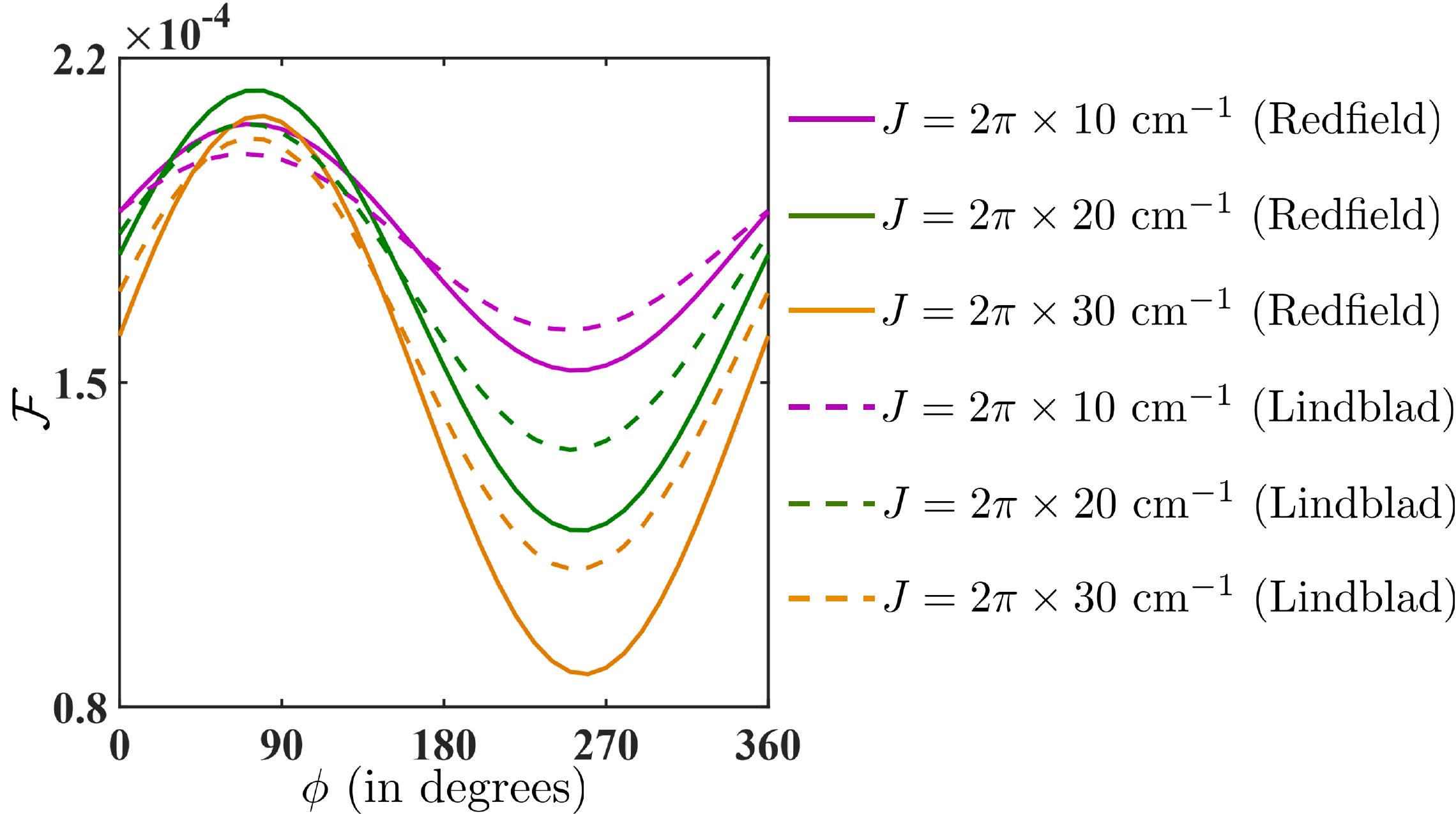}
\caption{Illustration of flux as a function of the relative phase between laser beams for different values of Coulomb couplings for Redfield and Lindblad approach. The parameters used in these simulations are provided in the main text.
}
\label{Redfield_lindblad}
\end{figure*}

Note that instead of the Redfield approach (explained above), we have also used the Lindblad approach to simulate system bath interactions in a reaction center connected donor-acceptor pair\cite{jung_energy_2020}. The comparative analysis between these approaches shows a quantitative difference in simulated flux [see Fig.\ref{Redfield_lindblad}], although the overall parameter dependence is similar.

\subsection{System radiation Liouvillian}
\label{subsec:System-radiation part}
In the case of the system-radiation interaction, excitation and de-excitation of the donor and acceptor states results in four elementary optical transition channels.
It is therefore convenient to label each interaction channel by a pair $(m,\sigma)$, where $m\in\{d,a\}$ denotes the molecular site and $\sigma\in\{\downarrow,\uparrow\}$ specifies the direction of the optical transition. With this, system and radiation bath operators can be defined as
\begin{align}
\hat S_{m,\downarrow}
&\equiv
\ket{g}\bra{m},
&
\hat S_{m,\uparrow}
&\equiv
\ket{m}\bra{g},
\label{radiation_system_operators}
\\
\hat\Gamma_{m,\downarrow}
&\equiv
\sum_{\mathbf{k},\lambda}
(g_{\textbf{k},\lambda}^\ast)^m
\hat r_{\mathbf{k}\lambda}^{\dagger},
&
\hat\Gamma_{m,\uparrow}
&\equiv
\sum_{\mathbf{k},\lambda}
(g_{\textbf{k},\lambda})^m
\hat r_{\mathbf{k}\lambda}.
\label{radiation_bath_operators}
\end{align}
Here, $\hat S_{m,\downarrow}$ describes de-excitation of site $m$, accompanied by photon creation through $\hat\Gamma_{m,\downarrow}$, whereas $\hat S_{m,\uparrow}$ describes optical excitation of site $m$, accompanied by photon annihilation through $\hat\Gamma_{m,\uparrow}$. Now, the system-radiation Liouvillian can be expressed as

%In terms of these physically resolved transition channels, the
%second-order system--radiation generator can be written as
\begin{align}
\mathcal L_{SR}
={}&
-\sum_{m,n\in\{d,a\}}
\sum_{\sigma,\sigma'\in\{\downarrow,\uparrow\}}
\int_{0}^{\infty}d\tau\,
\Big[
\hat S_{m,\sigma}
\hat S_{n,\sigma'}(\tau)\rho\,
\big\langle
\hat\Gamma_{m,\sigma}
\hat\Gamma_{n,\sigma'}(\tau)
\big\rangle_R
\nonumber\\
&\hspace{3.2cm}
-\hat S_{m,\sigma}\rho
\hat S_{n,\sigma'}(\tau)\,
\big\langle
\hat\Gamma_{n,\sigma'}(\tau)
\hat\Gamma_{m,\sigma}
\big\rangle_R
\nonumber\\
&\hspace{3.2cm}
-\hat S_{n,\sigma'}(\tau)\rho
\hat S_{m,\sigma}\,
\big\langle
\hat\Gamma_{m,\sigma}
\hat\Gamma_{n,\sigma'}(\tau)
\big\rangle_R
\nonumber\\
&\hspace{3.2cm}
+\rho\hat S_{n,\sigma'}(\tau)
\hat S_{m,\sigma}\,
\big\langle
\hat\Gamma_{n,\sigma'}(\tau)
\hat\Gamma_{m,\sigma}
\big\rangle_R
\Big].
\label{Sys_rad_1}
\end{align}

The sums over $m,n\in\{d,a\}$ and $\sigma,\sigma'\in\{\downarrow,\uparrow\}$ generate $16$ ordered pairs of optical transition channels. Since $\langle r_{\textbf{k},\lambda}^\dagger r_{\textbf{k},\lambda}^\dagger \rangle = \langle r_{\textbf{k},\lambda} r_{\textbf{k},\lambda} \rangle \equiv 0$, $8$  out of $16$ combinations vanish. Each channel that survives contains four operator products in the fully expanded form of Eq.~\eqref{Sys_rad_1}. The corresponding radiation correlation functions determine whether a particular contribution describes spontaneous emission, stimulated emission, photon absorption, or interference between the donor and acceptor optical transitions.

Again by using Baker-Campbell-Hausdorff formula, the time dependence of operators simplifies to
\begin{align*}
\ket{d}\bra{g}(t) &= e^{i \frac{\omega_d + \omega_a}{2} t} \left[ (\cos(\omega_e t) + i \frac{\omega_d - \omega_a}{2\omega_e} \sin(\omega_e t)) \ket{d}\bra{g} + i \frac{J}{\omega_e} \sin(\omega_e t) \ket{a}\bra{g} \right], \\
\ket{a}\bra{g}(t) &= e^{i \frac{\omega_d + \omega_a}{2} t} \left[ i \frac{J}{\omega_e} \sin(\omega_e t) \ket{d}\bra{g} + (\cos(\omega_e t) - i \frac{\omega_d - \omega_a}{2\omega_e} \sin(\omega_e t)) \ket{a}\bra{g} \right].
\end{align*}
where $\omega_e = \sqrt{\left(\frac{\omega_d - \omega_a}{2}\right)^2 + J^2}$. 
Similarly, for the radiation part, the time dependence can be written as $\hat\Gamma_{d,\downarrow}(t) = \sum_{\textbf{k},\lambda} (g_{\textbf{k},\lambda}^\ast)^d r_{\textbf{k},\lambda}^\dagger e^{i\omega_k t}$, $\hat\Gamma_{d,\uparrow}(t) = \sum_{\textbf{k},\lambda} (g_{\textbf{k},\lambda})^d r_{\textbf{k},\lambda} e^{-i\omega_k t}$, $\hat\Gamma_{a,\downarrow}(t) = \sum_{\textbf{k},\lambda} (g_{\textbf{k},\lambda}^\ast)^a r_{\textbf{k},\lambda}^\dagger e^{i\omega_k t}$, $\hat\Gamma_{a,\uparrow}(t) = \sum_{\textbf{k},\lambda} (g_{\textbf{k},\lambda})^a r_{\textbf{k},\lambda} e^{-i\omega_k t}$. With this, Eq.(\ref{Sys_rad_1}) simplifies to
\begin{align}
\mathcal L_{SR}[\rho]
={}&
\sum_{m\in\{d,a\}}
\gamma_{m,m}
\Big[
(\bar n_m+1)\,
\mathcal D[\hat S_{m,\downarrow}]\rho
+
\bar n_m\,
\mathcal D[\hat S_{m,\uparrow}]\rho
\Big]
\nonumber\\[0.3em]
&+
\frac{1}{2}
\Big[
\gamma_{d,a}(\bar n_a+1)
+
\gamma_{a,d}(\bar n_d+1)
\Big]
\Bigg[
\hat S_{d,\downarrow}\rho\hat S_{a,\uparrow}
+
\hat S_{a,\downarrow}\rho\hat S_{d,\uparrow}
\nonumber\\
&\hspace{4.3cm}
-
\frac{1}{2}
\left\{
\hat S_{d,\uparrow}\hat S_{a,\downarrow}
+
\hat S_{a,\uparrow}\hat S_{d,\downarrow},
\rho
\right\}
\Bigg]
\nonumber\\[0.3em]
&-
\frac{1}{4}
\Big[
\gamma_{d,a}(\bar n_a+1)
-
\gamma_{a,d}(\bar n_d+1)
\Big]
\left[
\hat S_{d,\uparrow}\hat S_{a,\downarrow}
-
\hat S_{a,\uparrow}\hat S_{d,\downarrow},
\rho
\right]
\nonumber\\[0.3em]
&+
\frac{1}{2}
\Big(
\gamma_{d,a}\bar n_a
+
\gamma_{a,d}\bar n_d
\Big)
\Big(
\hat S_{d,\uparrow}\rho\hat S_{a,\downarrow}
+
\hat S_{a,\uparrow}\rho\hat S_{d,\downarrow}
\Big).
\label{Sys_rad_2}
\end{align}

Here, $\mathcal D[\hat L]\rho \equiv \hat L\rho\hat L^\dagger -\frac{1}{2}\left\{\hat L^\dagger\hat L,\rho\right\}$. 
Moreover, the spontaneous decay rates can be expressed as \cite{Carmichael}
%\begin{widetext}
\begin{align}
\gamma_{d,d(a,a)}\bar{n}_{d(a)} \equiv & 2\pi\sum_\lambda \int d^3\textbf{k} \mathcal{G}(\textbf{k}) (\omega_k/2\hbar\varepsilon_0V) (\hat{e}_{\textbf{k},\lambda}\cdot\vec{d}_{gd(ga)})(\hat{e}_{\textbf{k},\lambda}\cdot\vec{d}_{gd(ga)})\langle r_{\textbf{k},\lambda}^\dagger r_{\textbf{k},\lambda} \rangle \delta(\omega_k - \omega_{d(a)}), \nonumber\\
\gamma_{d,a}\bar{n}_a \equiv & 2\pi \textbf{p} \sum_\lambda \int d^3\textbf{k} \mathcal{G}(\textbf{k}) (\omega_k/2\hbar\varepsilon_0V) (\hat{e}_{\textbf{k},\lambda}\cdot\vec{d}_{gd})(\hat{e}_{\textbf{k},\lambda}\cdot\vec{d}_{ga}) \langle r_{\textbf{k},\lambda}^\dagger r_{\textbf{k},\lambda} \rangle \delta(\omega_k - \omega_{a}).
\end{align}  
%\end{widetext}
Here, $\textbf{p} = \cos\theta_{da}$ (with $\theta_{da}$ as the angle between donor and acceptor dipoles) is an alignment factor \cite{kozlov_inducing_2006}. 
The first line of Eq.(~\eqref{Sys_rad_2}) describes independent radiative transitions on the donor and acceptor. In particular, $\gamma_{mm}(\bar{n}_m+1)$ contains both spontaneous and stimulated emission, whereas $\gamma_{mm}\bar{n}_m$ describes photon absorption. The remaining terms originate from donor-acceptor cross correlations in the radiation field. They represent collective radiative decay, collective excitation, and interference between the two optical transition pathways.
For further simplification, following Scully et al. \cite{kozlov_inducing_2006}, consider $\bar{n}_d = \bar{n}_a = \bar{n}$ and introduce $\gamma_{m,n} = \textbf{p}_{m,n}\sqrt{\gamma_m \gamma_n}$ (such that $\textbf{p}_{m,n} = 1 $ when $m=n$ and $\textbf{p}_{m,n} = \textbf{p}$ when $m \neq n$) to describe incoherent pumping rate as $r_m = \sqrt{\gamma_m \bar{n}}$. The simplified form of the system-radiation field is then
\begin{align}
\mathcal L_{SR}
={}&
\sum_{m\in\{d,a\}}
\Big[
(r_m+\gamma_m)\,
\mathcal D[\hat S_{m,\downarrow}]\rho
+r_m\,
\mathcal D[\hat S_{m,\uparrow}]\rho
\Big]
\nonumber\\[0.35em]
&+
\textbf{p}\left(\sqrt{r_d r_a}+\sqrt{\gamma_d\gamma_a}\right)
\Big[
\mathcal D[\hat S_{d,\downarrow},\hat S_{a,\downarrow}]\rho
+
\mathcal D[\hat S_{a,\downarrow},\hat S_{d,\downarrow}]\rho
\Big]
\nonumber\\[0.35em]
&+
\textbf{p}\sqrt{r_d r_a}
\Big[
\mathcal D[\hat S_{d,\uparrow},\hat S_{a,\uparrow}]\rho
+
\mathcal D[\hat S_{a,\uparrow},\hat S_{d,\uparrow}]\rho
\Big].
\label{Sys_rad_3}
%\end{empheq}
\end{align}
Here, $\mathcal D[\hat A,\hat B]\rho \equiv \hat A\rho\hat B^{\dagger} -\frac{1}{2}\anticommutator{\hat B^{\dagger}\hat A}{\rho}$. Moreover, $\gamma_{m,m}$ is replaced by $\gamma_m$.

\subsection{Irreversible contributions}
\label{subsec:System-irre part}
Apart from the spontaneous decay, to account for the other channels of exciton recombination an irreversible contribution with rate $\Gamma_{e}$ is considered in the master equation\cite{jung_energy_2020}, i.e.,
\begin{align}
\mathcal L_e
&=
2\Gamma_e
\sum_{m\in\{d,a\}}
\mathcal D[\hat S_{m,\downarrow}]\rho,
\label{Eq_recomb}
\end{align}
In a similar manner, by using $\hat S_t \equiv \ket{r}\bra{a},$ migration of excitation from the acceptor site to the reaction center is considered as follows \cite{jung_energy_2020}
\begin{align}
\mathcal L_t
&=
2\Gamma_t\,
\mathcal D[\hat S_t]\rho,
\label{Eq_trap}
\end{align}
where, $\Gamma_{t}$ is the trapping rate for transition $\ket{a}\rightarrow\ket{r}$. In this process of consistent injection and migration of excitation in a donor-acceptor pair connected to the reaction center, for completeness of a light-reaction cycle (in an oversimplified manner), an additional channel from $\ket{r}$ to $\ket{g}$ (i.e., $\hat S_r\equiv \ket{g}\bra{r}$) is included in the simulation (with rate $\Gamma_{r}$) as 
\begin{align}
\mathcal L_r
&=
2\Gamma_r\,
\mathcal D[\hat S_r]\rho.
\label{Eq_recycle}
\end{align}
%\end{widetext}

\section{Analytical form of the NESS donor-acceptor flux}\label{subsec:NESS}
Note that exact expressions are lengthy and complicated.
However, for a specific phase relation, e.g., $\phi = \phi_d - \phi_a = 0^\circ$, a relatively simplified case can be considered with perfect resonance without phonon bath interactions such that by assuming $\Omega_d = \Omega_a = \Omega$, $r_d = r_a = r$, $\Gamma_{t} = \Gamma_{r} = \Gamma$, $\gamma_d = \gamma_a = \gamma_s$, $\Gamma_{e} = 0$, $\textbf{p} = 1$, the NESS donor-acceptor flux $\mathcal{F}$ simplifies to $\mathcal{F} = \mathcal{N}_e/\mathcal{D}_e$,
where numerator can be expressed as

\begin{equation}
\mathcal{N}_e
=
4\Gamma J^2
\left[
8\Omega^4 \mathcal{X}_1
-2\Omega^2 \mathcal{X}_2
-r\left(4J^4+4J^2\mathcal{X}_3+\mathcal{X}_4^2\right)
\right].
\label{eq:numerator}
\end{equation}
Such that
\begin{align*}
\mathcal{X}_1
&\equiv
\Gamma+2(r+\gs),
\\
\mathcal{X}_2
&\equiv
J^2(4\Gamma+6r+8\gs)
+r(2\Gamma+3\gs)(3\Gamma+2\gs)
+r^2(11\Gamma+6\gs)
+2\gs(\Gamma+\gs)^2,
\\
\mathcal{X}_3
&\equiv
5r^2+r(3\Gamma+4\gs)+\gs(\Gamma+\gs)\\
\mathcal{X}_4
&\equiv
\Gamma(3r+\gs)+2r(2r+\gs).
\end{align*}

Similarly, denominator can be written as

\begin{equation}
\mathcal{D}_e
=
\mathcal{Y}_1+2\Omega^2\mathcal{Y}_2+8\Omega^4\mathcal{Y}_3+80\Omega^6\mathcal{Y}_4.
\label{eq:denominator}
\end{equation}
Such that
\begin{align*}
\mathcal{Y}_1
={}&
\Big[
4J^4
+4J^2\left(
3\Gamma r+5r^2+4r\gs+\Gamma\gs+\gs^2
\right)
+A^2
\Big]
\nonumber\\
&\times
\Big[
4J^2(\Gamma+5r+2\gs)
+(\Gamma+r+\gs)
\left(
5r^2+r(4\Gamma+7\gs)+2\gs(\Gamma+\gs)
\right)
\Big],
\\[1ex]
\mathcal{Y}_2
={}&
8J^4(3\Gamma+10r+6\gs)
\nonumber\\
&+
2J^2
\Big[
8\Gamma^3
+\Gamma r(55r+17\gs)
+6\Gamma^2(7r+3\gs)
-2\left(
20r^3+53r^2\gs+28r\gs^2+3\gs^3
\right)
\Big]
\nonumber\\
&+
\mathcal{X}_4
\Big[
50r^3
+r\left(32\Gamma^2+81\Gamma\gs+54\gs^2\right)
+r^2(71\Gamma+96\gs)
+4(\Gamma+\gs)
\left(2\Gamma^2+3\Gamma\gs+2\gs^2\right)
\Big],
\\
\mathcal{Y}_3
={}&
4\Gamma^3
-2J^2(9\Gamma+35r+18\gs)
+
\Gamma\left(90r^2+84r\gs+17\gs^2\right)
+\Gamma^2(41r+18\gs)
\nonumber\\
&+
2(2r+\gs)
\left(20r^2+19r\gs+2\gs^2\right),
\\[1ex]
\mathcal{Y}_4
={}&
\Gamma+4r+2\gs.
\end{align*}

Now, the flux in the presence of incoherent pumping only (that is, in the absence of coherent pumping) is
\begin{align*}
\mathcal{F}\mid_{\Omega = 0} &= \frac{4 \Gamma J^2 r}{4 J^2 \left(\Gamma+5 r\right)+r \left(4 \Gamma^2+9 \Gamma r+5 r^2\right)}.
\end{align*}
Usually, photosynthetic complexes operate with $r/J(\Gamma) \lesssim 10^{-2} $, it implies flux is a linear function incoherent pumping, i.e., $\mathcal{F}\mid_{\Omega = 0} \sim r$.
On the contrary, the presence of coherent pumping only (that is, in the absence of incoherent pumping) leads to
\begin{align}
\mathcal{F}\mid_{r = 0} &= \frac{2 \Gamma \Omega^2 J^2 \left(J^2-\Omega^2\right)}{5 \Omega^6+J^6+\Omega^4 \left(2 \Gamma^2-9 J^2\right)+ d_2 \Omega^2}, \nonumber\\
%d_2 &= \Omega^2 \left(2 \Gamma^2 J^2+3 \left(J\right)^4\right).
\label{coh_pr}
\end{align}
here, $d_2 =  2 \Gamma^2 J^2+3 \left(J\right)^4$.
With $0\lesssim\Omega\leqslant5 $, it indicates the following three cases: 
\begin{enumerate}
\item[$(\rNum{1})$] When Coulomb coupling is relatively strong such that $\Omega/J (\Gamma) \lesssim 10^{-2} $, flux follows $\mathcal{F}\mid_{r = 0} \sim \Omega^2 \left(J\right)^{-1}$.

\item[$(\rNum{2})$] When Coulomb coupling between dipoles is weak such that $\Omega \rightarrow J$, the $\mathcal{F}\mid_{r=0} \rightarrow 0$. 

\item[$(\rNum{3})$] For intermediate Coulomb coupling, i.e., $\Omega/J(\Gamma) \sim 10^{-1}$, the flux is $\mathcal{F}\mid_{r=0} \sim 10^{-1} \Omega$. %In implies that %with $0.1\lesssim\Omega\leqslant5 $, 
%a significant flux can be injected by coherent field compared to incoherent pumping.
\end{enumerate}
This implies that in extreme cases $(\rNum{1})$ and $(\rNum{2})$, the coherent field injects less flux than for intermediate Coulomb coupling. 
It indicates the existence of a maxima of the flux for each value of $\Omega$ as a function of $J$ in the presence of the coherent field.

%%%%%%%%%%%%%%%%%%%%%%%%%%%%%%%%%%%%%%%%%%%%%%%%%%%%%%%%%%%%%%%%%

\bibliography{Paper_flux_cont}

\end{document}